
\input harvmac
\noblackbox
\def \eq#1 {\eqno {(#1)}}
\def \cB {{\cal B}}
\def \cA {{\cal A}}

\def \ra {\rightarrow}
\def\k{\kappa}
\def\r{\rho}
\def\a{\alpha}
\def\b{\beta}

\def\d{\delta}

\def\e{\epsilon}

\def\p{\phi}

\def\m{\mu}
\def\n{\nu}

\def\l{\lambda}

\def\s{\sigma}

\def\cA{{\cal A}}
\def \sm {$\s$-model\ }

\def \bd {\bar \del}

\def \ha {{1\over 2}}

\def \ov {\over}

\def\const{{\rm const}}
\def \p {\phi}
\def \vp {\varphi}

\def \bd {\bar \del}

\def \ra {\rightarrow}

\def \na {\nabla }

\def \a {\alpha}
\def \b {\beta}

\def \ln {{\rm \ ln \  }}
\def \det {{\ \rm det \ }}

\def \l {\lambda}
\def \p {\phi}

\def \m {\mu }
\def \n {\nu}

\def \r {\rho}
\def \k {\kappa }
\def \d {\delta}

\def \s {\sigma}

\def \fourth {{\textstyle{1\over 4}}}

\def \e#1 {{{\rm e}^{#1}}}
\def \const {{\rm const }}

\def \eq#1 {\eqno {(#1)}}
\def \sm {$\s$-model\ }

\def \bd  {{ \bar \del }}

\def \bd  { \bar \del }


\def \p {\phi}

\def \s {\sigma}

\def \r {\rho}
\def \d {\delta}
\def \l {\lambda}
\def \m {\mu}

\def \n {\nu}

\def \fourth {{1\over 4}}

\def \e#1 {{{\rm e}^{#1}}}
\def \const {{\rm const }}\def \vp {\varphi}

\def \m {\mu}

\def \ra {\rightarrow}

\def \const {{\rm const} }

\def \eq#1 {\eqno{(#1)}}
\def \e {\rm e}
\def \ra {\rightarrow }
\def \e#1 {{\rm e}^{#1}}

\def \ln { {\rm ln } }

\def \l {\lambda}
\def \p {\phi}
\def \vp {\varphi}

\def \r {\rho}

\def\({\left (}
\def\){\right )}
\def\[{\left [}
\def\]{\right ]}

\def\bd {{\bar \del}}\def \ra {\rightarrow}
\def \vp {\varphi}

\def \eq#1 {\eqno{(#1)}}

\def \a {\alpha}

\def \b {\beta}
\def \k {\kappa}

\def \p {\phi}

\def \s {\sigma}
\def \r {\rho}
\def \d {\delta}
\def \l {\lambda}
\def \m {\mu}

\def \n {\nu}

\def \fourth {{1\over 4}}

\def \e#1 {{\rm e}^{#1}}
\def \const {{\rm const }}

\def \vp {\varphi}

\def \ha { { 1\over 2 }}

\def \ov {\over}

\def \sm  { sigma model\ }

\def\np {  Nucl. Phys. }
\def \pl { Phys. Lett. }
\def \mpl { Mod. Phys. Lett. }
\def \prl { Phys. Rev. Lett. }
\def \pr  { Phys. Rev. }

\def \ijmp { Int. J. Mod. Phys. }
\def \cqg {Class. Quant. Grav.}

\lref \hhtt{ J. Horne, G. Horowitz and A. Steif, \jnl \prl,  68, 568, 1992.}

 \lref \busc { T.H. Buscher, \pl B194(1987)59 ; \pl B201(1988)466.}
\lref \pan  { J. Panvel, \pl B284(1992)50. }

\lref \mye {  R. Myers, \pl B199(1987)371;
    I. Antoniadis, C. Bachas, J. Ellis, D. Nanopoulos,
\pl B211(1988)393;
 \np B328(1989)115. }

\lref  \horne { J.H. Horne and G.T. Horowitz, \np B368(1992)444.}

\lref \koki { K. Kounnas and  E. Kiritsis, preprint CERN-TH.7059/93;
hep-th/9310202. }

\lref \tsmpl {A.A. Tseytlin, \mpl A6(1991)1721.}

\lref \tsbh {A.A. Tseytlin,  preprint CERN-TH.6970/93; hep-th/9308042.}

\lref \nsw {  K.S. Narain, M.H. Sarmadi and E. Witten, \np B279(1987)369. }

\lref \tsdu { A.A. Tseytlin, \pl B242(1990)163; \np B350(1991)395.  }


\lref \wi { E. Witten, unpublished (1991). }
\lref \kik {  K.~Kikkawa and M.~Yamanaka,  \jnl \pl, B149, 357, 1984;
N.~Sakai and I.~Senda, {{\it  Progr. Theor. Phys.}}
{{\bf 75}} (1986) 692;
M.~B.~Green,
 J.~H.~Schwarz and L.~Brink, {{\it Nucl.Phys}}.  {{\bf B198}} (1982) 474. }
\lref\nair{ V.~Nair, A.~Shapere, A.~Strominger and F.~Wilczek,
{{\it Nucl. Phys}}. {{\bf B287}} (1987) 402.
}

\lref \vaf {R. Brandenberger and C. Vafa, \jnl \np, B316,  391, 1988; A.
Tseytlin and C. Vafa, \jnl \np,  B372, 443, 1992.}
\lref \por{ A. Giveon, M. Porrati and E. Rabinovici, preprint RI-1-94,
hep-th/9401139. }
\lref \hht{G. Horowitz and A. Steif, \jnl \pl,  B258, 91, 1991.}

\lref \klits{  C. Klim\v c\'\i k  and A.A. Tseytlin, unpublished (1994). }
\def \lr { \lref}

\gdef \jnl#1, #2, #3, 1#4#5#6{ {\sl #1~}{\bf #2} (1#4#5#6) #3}

\lr \mans { P. Mansfield and J. Miramontes, \jnl \pl, B199, 224, 1988;
A. Tseytlin, \jnl \pl, B208, 228, 1988; \jnl \pl, B223, 165, 1989.}

\lr \kalmor{R. Kallosh and A. Morozov,  \jnl \ijmp,  A3, 1943, 1988.}

\lr \ghrw{J. Gauntlett, J. Harvey, M. Robinson, and D. Waldram,
\jnl \np, B411, 461, 1994.}
\lr \garf{D. Garfinkle, \jnl \pr, D46, 4286, 1992.}

\lref \tspl {A. Tseytlin, \jnl \pl, B317, 559, 1993.}
\lref \tssfet { K. Sfetsos and A.  Tseytlin, \jnl  \pr, D49, 2933, 1994.}
\lref \klts {C. Klim\v c\'\i k  and A. Tseytlin, ``Exact four dimensional
string solutions and Toda-like sigma models from null-gauged
WZNW models",  preprint
 Imperial/TP/93-94/17, hep-th/9402120.}

\lr \sfexac {K. Sfetsos,  \jnl \np, B389, 424,  1993.}

\lr \tsmac{A. Tseytlin, \jnl \pl,  B251, 530, 1990.}

\lr \cakh{C. Callan and R. Khuri, \jnl \pl, B261, 363, 1991;
R. Khuri, \jnl \np, B403, 335, 1993.}
\lr \dgt{M. Duff, G. Gibbons and P. Townsend, ``Macroscopic superstrings
as interpolating solitons", DAMTP/R-93/5, hep-th/9405124.}

\lref \ger {A. Gerasimov, A. Morozov, M. Olshanetsky, A. Marshakov and S.
Shatashvili, \jnl \ijmp,
A5, 2495,  1990. }

\lr \hutow{C. Hull and P. Townsend, \jnl \np, B274, 349, 1986.}
\lr \mukh {S. Mukhi,    \jnl \pl,  B162, 345, 1985;
S. De Alwis, \jnl \pl, B164, 67, 1985. }

\lr \scherk { J. Scherk, \jnl \np, B31,  222, 1971;
   J. Scherk and J. Schwarz, \jnl \np, B81, 118,
1974;  T. Yoneya, \jnl {\it Progr. Theor. Phys.}, 51, 1907, 1974.}
 \lr \lov  {C. Lovelace,  \jnl \pl,  B135, 75, 1984;\jnl \np,  B273, 413,
1986.}
\lr \call{C. Callan, D. Friedan, E. Martinec and  M. Perry, \jnl \np, B262,
593, 1985.}
\lr \frts {E.  Fradkin  and A. Tseytlin, \jnl \pl, B158, 316, 1985;
\jnl \np, B261, 1, 1985.}
\lr \tsred  {A. Tseytlin, \jnl  \pl, B176, 92, 1986; \jnl  \np, B276, 391,
 1986.}
\lr \grwi   { D. Gross and E. Witten, \jnl \np, B277, 1, 1986.}

\lr \gps {S.  Giddings, J. Polchinski and A. Strominger, \jnl  \pr,  D48,
 5784, 1993. }

\lr \tsppl  {A. Tseytlin, \jnl   \pl,  B208, 221, 1988.}
\lr\rabi  {S. Elitzur, A. Forge and E. Rabinovici, \jnl \np, B359, 581, 1991;
 G. Mandal, A. Sengupta and S. Wadia, \jnl \mpl,  A6, 1685, 1991. }
 \lr \witt{ E. Witten, \jnl \pr, D44, 314, 1991. }
 \lr \dvv { R. Dijkgraaf, H. Verlinde and E. Verlinde, \jnl \np, B371,
269, 1992.}
\lr \hoho { J. Horne and G.  Horowitz, \jnl \np, B368, 444, 1992. }
\lr \horwel{G. Horowitz and D. Welch, \jnl \prl, 71, 328, 1993;
N. Kaloper,  \jnl \pr,  D48, 2598, 1993. }
\lr \host{ G. Horowitz and A. Steif,  \jnl \prl, 64, 260, 1990; \jnl \pr,
D42, 1950, 1990;  G. Horowitz, in: {\it
 Strings '90}, (eds. R Arnowitt et. al.)
 World Scientific (1991).}
\lr \busch {T.  Buscher, \jnl \pl, B194, 59, 1987; \jnl \pl,
 B201, 466, 1988.}
\lr \kallosh {E. Bergshoeff, I. Entrop, and R. Kallosh, ``Exact Duality in
String Effective Action", SU-ITP-93-37; hep-th/9401025.}

\lr \tsmpl {A. Tseytlin, \jnl  \mpl, A6, 1721, 1991.}
\lr \vene { }
\lr \kltspl { C. Klim\v c\'\i k and A. Tseytlin, \jnl \pl, B323, 305, 1994.}
\lr \shwts { A. Schwarz and A. Tseytlin, \jnl \np, B399, 691, 1993.}
\lr \callnts { C. Callan and Z. Gan, \jnl  \np, B272, 647, 1986;  A. Tseytlin,
\jnl \pl,
B178, 34, 1986.}

\lr \desa{ H. de Vega and N. Sanchez, \jnl
\pr, D45, 2783, 1992; \jnl \cqg, 10, 2007, 1993.}
\lr \desas{ H. de Vega and N. Sanchez, \jnl
\pl, B244,  215, 1990.}
\lref \tsnul { A. Tseytlin, \jnl \np, B390, 153, 1993.}

\lref \dunu { G. Horowitz and A. Steif, \pl B250 (1990) 49;
 E. Smith and J. Polchinski, \pl B263 (1991) 59. }

\lr \gauged {I. Bars and K. Sfetsos, \jnl  \mpl, A7, 1091, 1992;
 P. Ginsparg and F. Quevedo, \jnl \np, B385, 527, 1992. }
\lr \bsfet {I. Bars and K. Sfetsos, \jnl \pr, D46, 4510, 1992; \jnl \pr,
 D48, 844, 1993. }
\lr \tsnp{ A. Tseytlin, \jnl \np, B399, 601, 1993;  \jnl \np, B411, 509, 1994.}
\lr \gibb{A. Dabholkar, G. Gibbons, J. Harvey and F. Ruiz Ruiz, \jnl \np, B340,
33, 1990.}
\lr \hhs{
G. Horowitz, in:  {\it  String Theory and Quantum Gravity '92,  Proc. of the
1992 Trieste Spring School}, ed.  J. Harvey et al. (World Scientific,
Singapore, 1993); hep-th/9210119.}

\lr \horstr{G. Horowitz and A. Strominger, \jnl \np, B360, 197, 1991.}

\lr \givkir {A.  Giveon and E. Kiritsis, \jnl \np, B411, 487, 1994.  }

\lr \jac{I. Jack and D. Jones, \jnl \pl, B200, 453, 1988.}
\lr \metts{R. Metsaev and A. Tseytlin, \jnl \np,  B293, 385, 1987.  }
\lr \horv{ P. Horava, \jnl \pl,
B278, 101, 1992.}

\lref \FT {E.S. Fradkin and A.A. Tseytlin, Phys.Lett. B158(1985)316; Nucl.Phys.
B261(1985)1. }
\lref \love {C. Lovelace, \pl B135(1984)75; \np B273(1986)413.}
\lref \sch {J. Scherk and J.H. Schwarz, \np B81(1974)118.}

\lref \por{ A. Giveon, M. Porrati and E. Rabinovici, preprint RI-1-94,
hep-th/9401139. }
\lref \klits{  C. Klim\v c\'\i k  and A.A. Tseytlin, unpublished (1994). }

\lr \horts { G. Horowitz and A. Tseytlin, ``On exact solutions and
singularities in string theory", preprint  Imperial/TP/93-94/38;
hep-th/9406067.}

\lr \hrts { G. Horowitz and A. Tseytlin, ``A new class of exact solutions
in string theory",   preprint  Imperial/TP/93-94/54, UCSBTH-94-31,
hep-th/9409021.}

\lr \tsmac{A. Tseytlin, \jnl \pl,  B251, 530, 1990.}

\lr \tspa { A. Tseytlin, ``Exact  string solutions
and duality", to appear in:  {\it Proceedings of the 2nd Journ\'ee Cosmologie,
 Observatoire de Paris, June 2-4, 1994}, ed. H. De Vega and N. Sanchez (World
Scientific, Singapore);  hep-th/9407099.}

\lr\bergsh { E. Bergshoeff, R. Kallosh and T. Ort\'in, \jnl \pr,  D47, 5444,
1993.}

\lr\sen{A. Sen, \jnl \np, B388, 457, 1992. }
\lr \garf{D. Garfinkle, \jnl \pr, D46, 4286, 1992.}
\lr \wald {D. Waldram, \jnl \pr, D47, 2528, 1993.}

\lr\gibbb{G. Gibbons, talks at LMS  Symposium
 ``Quantum concepts in time and space", Durham, July 1994  and  International
Congress of Mathematical Physics, Paris, July 1994. }

\lr \born { M. Born, {\sl Proc. Roy. Soc. } {\bf A143} (1934) 410;
 M. Born and L. Infeld, {\sl Proc. Roy. Soc. } {\bf A144} (1934) 425.}

\lr \hrt  { G. Horowitz and A. Tseytlin, ``Extremal black holes as exact string
solutions", preprint  Imperial/TP/93-94/51, UCSBTH-94-24, hep-th/9408040.}

\lr \gib { G. Gibbons, \jnl \np, B207, 337, 1982. }

\lr \gim { G. Gibbons and K. Maeda,  \jnl \np, B298, 741, 1988. }
\lr\gar { D. Garfinkle, G. Horowitz and A. Strominger, \jnl \pr,  D43, 3140,
1991; {\bf D45} (1992) 3888(E). }
 \lr \witt{ E. Witten, \jnl \pr, D44, 314, 1991. }

\lr \guv { R. G\"uven, \jnl \pl, B191, 275, 1987;
 D. Amati and C. Klim\v c\'\i k,
\jnl \pl, B219, 443, 1989; G. Horowitz and A. Steif,  \jnl \prl, 64, 260,
1990.}

\lr\kall { R. Kallosh, D. Kastor, T. Ort\'in   and T. Torma, ``Supersymmetry
and stationary solutions in dilaton-axion gravity",
SU-ITP-94-12, hep-th/9406059. }
\lr\joh { C. Johnson and R. Myers, ``Taub--NUT Dyons in Heterotic String
Theory",
IASSNS-HEP-94/50, McGill/94-28, hep-th/9406069. }

\lr\jons { C. Jonson and R.  Myers, ``Taub-NUT dyons in heterotic string
theory",
 IASSNS-HEP-94/50, hep-th/9406069. }

\lr \jon {C. Johnson,  ``Exact models of extremal dyonic 4D black hole
solutions of heterotic string theory", IASSNS-HEP-94/20,
hep-th/9403192.}

\lr \gurs{M. G\"urses, \jnl \pr, D46, 2522, 1992.}

\lr\bergsh { E. Bergshoeff, R. Kallosh and T. Ort\'in, \jnl \pr,  D47, 5444,
1993; E. Bergshoeff, I. Entrop and R. Kallosh,
\jnl \pr, D49, 6663, 1994.}
\lr \wald {D. Waldram, \jnl \pr, D47, 2528, 1993.}

\lr \callan { C. Callan, R. Myers and M. Perry, \jnl \np, B311, 673, 1988;
R. Myers, \jnl \np, B289, 701, 1987.}

\lr\rot{J.~Horne and G.~Horowitz, \jnl \pr, D46, 1340, 1992;
A. Sen, \jnl \prl, 69, 1006, 1992.}
\lr \ghrw{J. Gauntlett, J. Harvey, M. Robinson and D. Waldram,
\jnl \np, B411, 461, 1994.}
\lr \garf{D. Garfinkle, \jnl \pr, D46, 4286, 1992.}

\lr\tser { A. Tseytlin, ``String cosmology and dilaton",  in: {\it
String Quantum Gravity  and Physics at the Planck scale,
Proceedings of the 1992 Erice
workshop}, ed. N. Sanchez (World
Scientific, Singapore, 1993) p.202; hep-th/9206067; \jnl \ijmp, D1, 223, 1992,
hep-th/9203033. }
\lr \tscv{M. Cveti\v c  and A. Tseytlin, \jnl \np, B416, 137, 1994.}
\lr \poldam{ T. Damour and A. Polyakov, \jnl \np, B423, 532, 1994. }
\lr \harms{B. Harms and Y. Leblanc,
``Conjectures on nonlocal effects in string black holes", UAHEP-935,
hep-th/9307042.}
\lr \frtse{E. Fradkin and A. Tseytlin, \jnl \pl,  B163, 123, 1985. }

\lr \cakh{C. Callan and R. Khuri, \jnl \pl, B261, 363, 1991;
R. Khuri, \jnl \np, B403, 335, 1993.}
\lr \ghrw{J. Gauntlett, J. Harvey, M. Robinson and D. Waldram,
\jnl \np, B411, 461, 1994.}

\lr \chs {C. Callan, J. Harvey and A. Strominger, \jnl \np,  B359,
 611,  1991.}

\lr \khga {R. Khuri, \jnl \np, B387, 315, 1992;  \jnl \pl, B294,
325, 1992; J. Gauntlett, J. Harvey and J. Liu, \jnl \np, B409, 363, 1993.}

\lr \dukh {M. Duff, R.  Khuri,  R. Minasyan and J. Rahmfeld,
\jnl \np, B418, 195, 1994. }
\lr \susy  {R. Kallosh, A. Linde, T. Ort\'in, A. Peet and A. Van Proeyen,  \jnl
\pr, D46, 5278, 1992.}
\lr \gaun{J. Gauntlett, talk presented at the conference
``Quantum Aspects of Black Holes",
Santa Barbara, June 1993.}
\lr\gersh{D. Gershon, \jnl \pr, D49, 999, 1994.}
\lr\wil{C. Holzhey and F. Wilczek, \jnl \np, B380, 447, 1992.}
\lr\edw{E. Witten, \jnl \pl, B149, 351, 1984.}
\lr \andr{O. Andreev and A. Tseytlin, \jnl \np, B311, 205, 1988; \jnl \mpl, A3,
1349, 1988.}
\lr \kalor { R. Kallosh and T. Ort\'in, ``Exact $SU(2)\times U(1)$ stringy
black holes", SU-ITP-94-27, hep-th/9409060. }

\lr\markov{ M. Markov, {\sl Usp. Fiz. Nauk}, {\bf 37} (1994) 57;
V. Frolov, M. Markov and V. Mukhanov, \jnl \pr, D41, 383, 1990.}
\lr \caa{H. Dorn and H. Otto, {\sl Z. Phys.} {\bf C32} (1986) 599;
A. Abouelsaood, C. Callan, C. Nappi and S. Yost, \jnl \np, B280, 599, 1987.}


\baselineskip8pt
\Title{\vbox
{\baselineskip 6pt{\hbox{Imperial/TP/93-94/59  }}{\hbox
{ hep-th/9410008 }}{\hbox{ October 1994}}\bigskip\bigskip\bigskip
{\hbox{ Dedicated to the memory of M.A. Markov  }} } }
{\vbox{\centerline {  Black Holes  and Exact Solutions }
\medskip
\centerline {   in String Theory   }
 }}
\vskip 8 true pt
\centerline{   A.A. Tseytlin\footnote{$^{\star}$}{\baselineskip8pt
e-mail address: tseytlin@ic.ac.uk}\footnote{$^{\dagger}$}{\baselineskip5pt
On leave  from Lebedev  Physics
Institute, Moscow, Russia.} }

\smallskip\smallskip
\centerline {\it  Theoretical Physics Group, Blackett Laboratory}

\centerline {\it  Imperial College,  London SW7 2BZ, U.K. }
\bigskip\bigskip

\centerline {\bf Abstract}
\medskip
\baselineskip12pt
\noindent
We  review  some recent results about exact
classical solutions in string theory.
In particular,  we consider four dimensional  extremal electric  black holes
which are related via dimensional reduction to the exact five dimensional
fundamental string solutions. We also comment on the issue of  $\a'$
corrections to  non-extremal black holes.

\bigskip
\bigskip\bigskip
\bigskip
\bigskip\bigskip
\bigskip

\centerline{ To appear in: {\it Proceedings of the  International School of
Astrophysics ``D. Chalonge", }}
\centerline{{\it 3rd course: Current Topics in Astrofundamental Physics,} }
\centerline{\it  4-16 September 1994, Erice, ed. N. Sanchez   (World
Scientific, Singapore)}

\Date { }


\newsec{Introduction}
 String theory
is the only presently known model of a unified theory which includes gravity
  and at the same time is consistent at the quantum level.
This, in principle, enables one to study   its quantum-gravitational
implications for  cosmology, black hole physics,
 space-time singularities, etc.

The basic objects of string theory are (closed) strings
which, when first-quantised,  have a finite number of massless
(graviton or metric $G_{\m\n}$, antisymmetric tensor $B_{\m\n}$, dilaton $\p$,
vectors, other scalars and spinors)
 and an infinite number of massive ($M^2 \sim 1/\a' \sim M^2_{Planck}$)
particle-like excitations.
Expanded near the trivial flat  space vacuum the theory is parametrised by two
constants: $\a'\sim L_{Planck}^2$ and dimensionless string coupling $g= \exp
\p_0$.The low-energy (large scale)  classical string dynamics  is described
by an  effective  action \refs{\scherk,\frts} $$S=\int d^Dx \sqrt G \e{-2\p}
 [ R +  4 (\del \p)^2 + ... + \a' R^2 + ...]\ , $$
 which starts with the Einstein term and  contains contributions  of all orders
in $\a'$.
  In general, the dilaton $\p$ plays the role of the (logarithm of the)
effective string coupling \refs{\edw,\frts}.

One of the important problems in  string theory is  to find
exact black hole solutions in four dimensions.
While   the vacuum Schwarzschild solution  of Einstein equations is also
a leading-order   solution in string theory,  it is modified by $\a'$
corrections.
It is straightforward to find other  solutions to the leading
order string effective equations describing charged dilatonic black holes
\refs{\gib,\gim,\gar}, as well as  to compute  the first nontrivial
$\a'$ corrections to the
Schwarzschild  solution  \callan.

To be able to study the    Planck-scale
 structure of the theory
one should go beyond the  solutions of the leading-order
  equations, i.e. one should  look for
 solutions which are  exact
in $\a'$.  Though   the explicit form  of the  full string effective action
is not known, one of the remarkable properties of string theory is that  it is
possible to by-pass this complication
by using the equivalence \refs{\lov,\call} between  extrema of the effective
action and 2-dimensional conformal theories (CFT's). Knowing the corresponding
CFT is also crucial for an adequate ``stringy" interpretation of a solution.

{ Exact}  black hole
solutions have been constructed in two \refs{\witt}   and three \horwel\
dimensions starting with
 (gauged) WZW models (or coset CFT's).
 In addition,
using the result that the extremal limit of a black five-brane \horstr\
is an exact superstring solution \chs, one can  obtain an exact
five dimensional extreme black hole (which has a magnetic  charge associated
with the antisymmetric tensor field).
It has recently been shown that one can also
dimensionally reduce the exact solution
of  \chs\ down to four dimensions \khga\ to find  an
extreme magnetically charged black hole
\dukh.\foot{These  exact five  and four  dimensional black hole solutions
are asymptotically
flat.  In addition, there are  coset CFT or gauged WZW
 constructions of just the `throat' regions  of some
 four dimensional extreme magnetically charged dilatonic black holes
\refs{\gps,\jon} (the fact that the form of the throat solution is unchanged
by $\a'$
corrections was noticed earlier \gurs).
Also, one can use gauged  WZW models to construct axisymmetric black-hole type
solutions \refs{\horv,\gersh}.} It was also suggested \kalor\ that extreme
magnetic dilatonic black hole, when properly embedded into higher dimensional
theory (with an extra gauge field background)
 is an exact solution of heterotic string.

Below we shall discuss  the     exact  string solution  \hrt\
describing an
extremal  four dimensional electrically charged black hole.\foot{The
electrically charged solutions are
qualitatively
different  from their magnetic analogues. In particular, the extremal
magnetically charged
solutions found in four and five dimensions are  products of a time-like line
and
a euclidean solution.
The extremal electrically charged solution, on the other hand, has a nontrivial
time-like direction.}  This solution can be considered  \hrt\ as
four-dimensional
``image" of an exact five-dimensional solution describing
the field of a fundamental string  \refs{\gibb,\wald} which winds around a
compact dimension.

In Section 2 we shall review the structure of the string effective action and
suggest a possibility  that string  theory is ``asymptotically free", i.e. that
the effective string coupling  goes to zero at small distances and  thus there
is a sense in studying classical solutions.
In Section 3 we shall  consider the leading-order terms in the string effective
action and present the corresponding  dilatonic black hole solution
\refs{\gib,\gim,\gar}.  We shall also comment on how $\a'$ corrections  can
possibly modify non-extremal black hole solutions
using the analogy with the Born-Infeld  effective action of the open string
theory.  In Section 4 we shall  briefly
discuss two classes of exact classical string solutions: ``plane fronted waves"
and ``fundamental string" backgrounds. Their close relation to extremal
electric
black hole  solutions (implying the exactness of the latter)
 will be explained in Section 5.
 Section 6 contains   concluding remarks.

\newsec{String effective action and  ``asymptotic freedom"  }

In general,
the  low-energy effective action in string theory is  given by  a sum of
classical, quantum (string loop) and non-perturbative contributions
\eqn\eff{ S=S_{c} + S_{l} + S_n \ . }
Several leading terms in the (bosonic string) classical
effective action are  \refs{\scherk,\frts,\call,\metts}\foot{ The form of such
action is  not  unique:
a class of actions
related by   field redefinitions  which are local,  covariant,
 background-independent, power series in $\a'$ (depending  on dilaton
 only through its derivatives not to mix different orders of string loop
 perturbation theory)  will correspond to the same string $S$-matrix
\refs{\tsred,\grwi}.}
\eqn\effc{ S_c = \int d^{D}x \sqrt {\mathstrut G} \ {\rm e}^{- 2
\phi} \big\{  c_0   +
  \  [ \   R \  + 4 \na^2 \p - 4 (\na \p )^2 - {1\ov 12} (H_{\m\n\l})^2\  ]  }
 $$  +  {1 \over 4} \a'\  [  \ R_{\m\n\l\k} R^{\m\n\l\k} -\ha R^{\m\n\k\l}H_{\
\m\n}^\r
H_{\k\l\r} $$ $$
+ {1\ov 24} H_{\m\n\l}H^\n_{\ \r\a}H^{\r\s\l}H_\s^{\ \m\a}  - {1\ov 8}
(H_{\m\a\b}H_\n^{\ \a\b})^2\ ] +
O(\a'^2)   \big\} \  ,  $$
where $c_0=-2(D-26)/ 3\a' .$
The loop contribution is ultraviolet-finite (with $\sqrt{ \a'}$ playing the
role of an effective  cutoff). For example, in the superstring theory $S_l$
has
the following  structure
\eqn\loo{ S_l= \sum^\infty_{n=0}  \int d^{D}x \sqrt {\mathstrut G} \ {\rm
e}^{2n
\phi}  ( {1\ov \a'} R^4 + ... + {\rm non}-{\rm local\  terms}  )\ . }
In  field theory non-perturbative terms are usually expected
to  involve  factors like $\exp( -1/g^2)$ where $g$ is a coupling constant.
Since in string theory  the latter is related to the dilaton,
the non-perturbative contributions  will  generate non-trivial  terms in
$S$ (dilaton potential, etc.)
\eqn\non{ S_n = \int d^{D}x \sqrt {\mathstrut G} \ V(\p) + ...\ ,
\ \ \  V= {c_1\ov g^2(\p) } \exp [{-{ c_2\ov g^2(\p) } }]  + ...\ , \ \ \ \ \
g(\p) =\e{ \p} \ . }
The non-perturbative corrections should be responsible
for supersymmetry breaking and generation of  a mass for the dilaton.
Since the dilaton has gravitational-type couplings it  should get a mass to
avoid contradiction  with observations (for a discussion and references see
e.g.
\refs{\tser, \poldam}).

The existence of a (perturbatively) massless scalar  dilaton
(which also  determines the strength of effective string coupling)  is one of
the universal and  most important consequences of string theory.
While dilaton should be effectively massive  at macroscopic distances
or late stages of evolution of the Universe, it may be massless
at small scales or early times.
In fact, one can contemplate a possibility of {\it  ``asymptotic freedom" }
in string theory: it may be that $\e{\p} $ is very small at small
scales/times so that string theory is effectively classical in early Universe
(or at  small distances), i.e.
 both perturbative and non-perturbative  string corrections  are negligible
there.
Dilaton may   grow at ``intermediate" scales/times  eventually   making
non-perturbative corrections  to become relevant and to  generate a potential
(mass) for the dilaton.\foot{It should be noted that a possible existence of
asymptotic freedom
of gravitational interactions  (and  limiting value of curvature)
was  suggested in \markov\ and references there. This idea becomes
especially   attractive  and feasible in  the context of string theory
because here there  is an additional field (dilaton) that governs the strength
of (gravitational) interaction.}
  It  seems  that
 the proper  dilaton  boundary condition  for, e.g.,      black hole  or
cosmological
 solutions in string
theory
  should be  $\p\ra -\infty$  at  $r\to 0 $ or $t\ra 0$,  that is   small
string
 coupling at small scales.\foot{There are, in fact,
 classical string solutions (in particular, the exact fundamental string
solutions discussed below)
that have such behaviour (see also  \refs{\gim,\gibb, \hhs,\tscv,\horts,\hrt}).
For a review of cosmological solutions see \tser.}

This  ``asymptotic freedom"   suggestion implies   that
massless dilaton should play an important role in early Universe (for example,
 the issue of primordial black holes should be addressed
on the basis of classical string theory with a massless dilaton).
Assuming  that string coupling is small at small distances,
in what follows we shall ignore both perturbative and non-perturbative
contributions to the string effective action and
discuss the {\it classical} black hole solutions in string theory.
The condition that string coupling should vanish  at small scales will,  in
fact,  be fulfilled
for  exact  solutions we shall describe.

%
\newsec{Leading-order   dilatonic black hole solutions  and  $\a'$ corrections
}
The leading-order terms in \effc\ may contain also the  Yang-Mills gauge field
term which may  be of  heterotic or
Kaluza-Klein origin.  Here we shall assume the second possibility.
If one considers a bosonic string in a space with some compact isometric
directions one can  get extra vector fields from ``non-diagonal" components
of the metric and the antisymmetric tensor.

For example, if we start with  the
leading-order term in the $D=5$ bosonic string action
\eqn\act{ S_5 =   \int d^5 x \sqrt G \  \e{-2\p}   \ \{
   \   R \ + 4 (\del_\m \p )^2 - {1\ov 12} (H_{\m\n\l})^2\
  + O(\a')   \}  \  , }
  and assume that all the fields are independent of $x^5=y$,  we obtain the
four
  dimensional reduced action
\eqn\acttp{  S_4 = \int d^4 x \sqrt {\hat G }\  \e{-2\p + \s}    \ \{
  \   \hat R \ + 4 (\del_\m \p )^2 - 4 \del_\m \p \del^\m \s }
$$  - {1\ov 12} (\hat H_{\m\n\l})^2\  - \fourth \e{2\s} ({ \cal F}_{\m\n})^2
-\fourth  \e{- 2\s} (\cB_{\m\n})^2
  + O(\a')   \}  \  , $$
where  we have defined
\eqn\fgfg{ G_{55}\equiv  \e{2\s}   ,   \ \  {\cal F}_{\m\n} = 2\del_{[\m}
\cA_{\n]}  \ ,   \ \ \cB_{\m\n} = 2 \del_{[\m} \cB_{\n]}  \  ,
\ \ \  \cA_\m\equiv   G^{55}  G_{\m 5}\ ,  \ \ \cB_\m \equiv    B_{\m 5}\ ,  }
$$\hat G_{\m\n} \equiv  G_{\m\n} - G_{55}\cA_\m \cA_\n
\ , \ \ \ \
 \hat H_{\l\m\n} = 3\del_{[\l} B_{\m\n]} - 3 \cA_{[\l} \cB_{\m\n]}
\  .  $$
This action can be considered also, for example,  as a part of $N=4, D=4$
supergravity action    obtained by dimensional reduction from $D=10$
supergravity.
Setting $\vp = 2\p -  \s$
the action \acttp\ becomes \hrts
\eqn\acttq{  S_4 = \int d^4 x \sqrt {\hat G }\  \e{- \vp}    \ \{
  \   \hat R \ +  (\del_\m  \vp )^2 - (\del_\m \s)^2  }
$$  - {1\ov 12} (\hat H_{\m\n\l})^2\  - \fourth \e{2\s} ({ \cal F}_{\m\n})^2
-\fourth  \e{- 2\s} (\cB_{\m\n})^2
  + O(\a')   \}  \  .  $$
In
 the Einstein frame (where  there  is no dilaton factor
in front of $R$)  \acttq\ takes the form
\eqn\acttw{S_4 = \int d^4 x \sqrt {{\hat G_E}}  \ \{
  \   \hat R_E \  - \ha  (\del_\m  \vp)^2 - (\del_\m \s)^2 } $$
 - \ {1\ov 12} \e{ - 2\vp  }   (\hat H_{\m\n\l})^2\
- \fourth \e{- \vp  +  2\s } ({ \cal F}_{\m\n})^2
-\fourth  \e{- \vp -  2\s} (\cB_{\m\n})^2
  + O(\a')   \}  \  . $$
Thus, in general, the four dimensional theory contains two scalars, two
vectors,
and the antisymmetric tensor, as well as the metric.
In certain special
cases, the nontrivial part of the action \acttw\ can be expressed in terms of
only one scalar and one vector,
so that it becomes
\eqn\actt{S_4 = \int d^4 x \sqrt {{\hat G_E}}   \ \{
  \   \hat R_E \ - \ha (\del_\m \psi )^2 - \fourth \e{- a\psi} ({ \cal
F}_{\m\n})^2  + O(\a') \}  \  . }
For example, if one sets $\p=0$  and $H_{\m\n\l} = 0$ in the $D=5$ action,  or
equivalently $\vp=-\s, \ {\hat H}_{\m\n\l} =0=\cB_{\m\n}$  directly in
\acttw, one
obtains \actt\ with $\psi= - a  \s$ and $ a=\sqrt 3$. This is, of course,
the standard Kaluza-Klein reduction of the Einstein action. Another possibility
which we shall mostly consider below is to set $\s =0\
(G_{55}=1),\ \ \hat H_{\m\n\l} =0$ and either the two vector fields
proportional
to each other, or let one of them vanish. This case corresponds to \actt\
 with $\psi= 2\p $  and $a =1$.

This  action   admits, of course,  the standard Schwarzschild solution
($R_{\m\n}=0,
\psi=0,  { \cal
F}_{\m\n}=0$)  but not the Reissner-Nordstrom one
(because of non-trivial coupling to the gauge field the dilaton cannot remain
constant).
The dilaton-vector  coupling plays  a crucial role in modifying the properties
of  black hole  - type solutions (this coupling  becomes irrelevant only after
the dilaton is ``frozen" at a minimum of a non-perturbative potential).

The charged dilatonic black hole  solutions (which are string analogues
of Reissner-Nordstrom one) were found in \refs{\gib,\gim,\gar} (for a review
see \hhs). The electrically charged solution has the folowing form
($\psi=2\phi, a=1$)
\eqn\ele{ ds^2= \e{2\p}  ds^2_E\ , \ \
ds^2_E= - (1- {2M\ov r}) dt^2 + (1- {2M\ov r})\inv dr^2 + r^2 (1- {2Q^2 \ov  M
r})
d\Omega^2_2 \ , }$$  {\cal A}_{t} = -{Q\ov r} \ , \ \ \ \e{2\p} = 1- {2Q^2 \ov
M r} \ . $$
This metric  has the same causal  structure as Schwarzschild (e.g. there is no
inner horizon) and singularity at $r=0$.

Taking the extremal limit $Q=M$ and redefining $r\ra r+ 2M$
one finds that the solution simplifies to
\eqn\elee{ ds^2= - F^2(r) dt^2 +  dr^2 + r^2 d\Omega^2_2 \ , }$$  {\cal A}_{t}
=  \e{2\p} = F(r) = {( 1 + {2M \ov   r})\inv  } \ . $$
Note that the spatial part of the metric   \elee\ is flat (this is one of the
reasons
 why this solution has  an exact generalisation to all orders in $\a'$).
Also, $\e{\p} \to 0 $ at $r\to 0$  so that the string coupling
becomes weak near the singularity (it  turns out that
this  remarkable behaviour is not modified by $\a'$-corrections).

There are various indications that extreme black holes play a special role
in string theory, being gravitational analogues of solitons
 \refs{\gib,\gim,\wil,\hhs,\gibbb}: they cannot be created from non-extremal
ones by a classical process, their entropy is zero, solutions can be linearly
superposed,
are supersymmetric and stable when embedded into supergravity theory \susy,
etc.  A solitonic interpretation
of extreme  electric  dilatonic black holes  is  strengthened
by the results of \hrt\ discussed below:
(i) extreme electric dilatonic  $D=4$ black hole solution  can be  represented
as a dimensional reduction  of the $D=5$ fundamental string solution solution
of \refs{\gibb,\wald} which itself has a solitonic interpretation and is  an
exact classical string solution \refs{\horts,\hrts}, so that (ii)  extreme
electric black hole solution
is not modified by $\a'$ corrections.
The absence of $\a'$ corrections is related to the extremality property
of the solution, i.e.  to a special balance between the metric and the gauge
field
(or the metric and the antisymmetric tensor  in the case of $D=5$ fundamental
string solution).

At the same time, the situation with non-extremal black holes
is  much more complicated: they are modified
by $\a'$ corrections in  a non-trivial
 way.\foot{As was already mentioned above, there is a field redefinition
ambiguity in the effective action. When saying that a solution is not modified
by $\a'$ corrections we understand that
this is true  modulo a local field redefinition (i.e. in  a specific
``scheme"):  a  solution in a generic scheme
will be related to the leading-order one by a {\it local} field redefinition.
A solution is said to receive non-trivial $\a'$ corrections if there does not
exist a  scheme  where the exact solution is equal to  the leading-order one.}
  For example,  starting with the Schwarzschild solution
(with $\p=\const$) and solving the effective equations  corresponding to $S_c$
\effc\
(assuming the central charge deficit $c_0$ is cancelled by some internal
degrees of freedom)
\eqn\efff{R_{\m\n} +  2 D_\m D_\n \p + \ha \a' R_{\m\l\r\k}  R_{\n}^{\ \l\r\k}
+ O(\a'^2) =0\  , }
\eqn\efr{ - D^2 \p + 2(\del \p)^2 + {1\ov 8} \a'  R_{\m\l\r\k}  R^{\m \l\r\k} +
O(\a'^2)=0  \ , }
one finds that to  the leading order in $\a'$  the   metric and  dilaton  get
$O(1/ r^n)$ corrections \callan. Thus
the dilaton is no longer constant.
Interestingly,  $\e{\p} $  does decrease towards the center
of black hole.
 Since the  Hawking-Penrose  singularity condition ($R_{\m\n}l^\m l^\n\geq 0, \
l^2\leq 0$) may not   be satisfied for the full  $\a'$-dependent effective
equations,  one may expect that $\a'$-corrections, once summed up,  may
 eliminate the $r=0$ singularity  by modifying  the  Schwarzschild metric
$G_{00}= -1 + {2M\ov r}$ ($M=GM_0$, \ $G\sim L^2_{Planck}\sim \a'$),  say, in
the following way
  (see also \harms)
\eqn\gue{ G_{00}= -1 + {2M \ov r}\exp (- c{ \a' M\ov r^3})\ .  }
In fact, the
 elimination of  singularity (and horizon)  seems   possible  in principle
only for black holes  of Planckian (or smaller mass), $M\sim
\sqrt{\a'}$.\foot{If such  elimination  of the singularity takes place it may
imply
that   massive  string states do not form singular black holes.}
In fact, for large masses $M\gg \sqrt {\a'}$ the  curvature is very small near
the (original)
horizon  $r_*=2M$ and $\a'$ corrections should be irrelevant there (in fact,
for $M\gg \sqrt {\a'}$ the correction factor in \gue\ is  close to one
 near   $r_*=2M$). Under the horizon the coordinates $r$ and $t$
interchange their roles and  one should actually look for a condition
of the absence of singularity at $t=0$.\foot{I am grateful to V. Mukhanov for a
discussion of this point.}

A possibility   that higher order $\a'$ terms in the effective action
may eliminate  the  black hole  singularity is suugested by  analogy
with the open string theory. There the   derivative-independent part of the
effective action for abelian gauge field can be found exactly
and is given by the Born-Infeld action \frtse\ (see also \caa; for derivative
corrections see \andr)
\eqn\bor{ S= \int d^Dx \sqrt {\det( \delta_{\m\n } + 2\pi \a' {\cal F}_{\m\n})
} + O(\a'^5{\cal F}^2 \del {\cal F} \del {\cal F}) \ . }
While in the Maxwell
theory the field of a point-like charge is singular at the origin and its
energy is infinite, in the Born-Infeld theory  the field is regular at $r=0$
where the electric field takes its maximal value
and its total energy is finite \born. From the point of view of the
distribution of the electric field ($\rho_{eff}=div E/ 4\pi$) the source is no
longer point-like  but  has an effective  radius  $r_0 \sim \sqrt {\a'}$ (for
example, in $D=4$  $ \  E_r={\cal F}_{rt}= Q/\sqrt{r^4 + r_0^4} , \ \  r_0^2=
2\pi\a'  Q $).
Since both open and closed string  theories are non-local
with a characteristic scale  of $\sqrt {\a'}$,
one may expect that the Schwarzschild singularity is smeared in a similar way.
Unfortunately, an  analogue  of the Born-Infeld action is  hard to derive
 in the closed string case, i.e.  there are technical difficulties on the way
to verification of this conjecture.

\newsec{Exact string solutions: ``plane waves" and ``fundamental strings"}
Backgrounds with null Killing vectors seem to play  a special  role in string
theory. It was observed in \guv\ that
the ``plane-wave" spaces ($u,v$ are ``light-cone" coordinates:  $u, v=y\pm t $;
\ $  i=1,...,D-2$)
\eqn\gu{ ds^2 = dudv + K(x,u)du^2 + dx^idx_i  \ , }
 represent exact classical string solutions if $K$ is subject to the
leading-order equation $R_{\m\n}=0, $ i.e.
$ \del_i\del^i K =0 $  (for example, $K= f(u) (x_1^2 -x_2^2)$).
There exist generalisations
of this  exact solution to the presence of antisymmetric tensor and dilaton
(see \refs{\tsnul,\bergsh,\horts,\hrts} and refs. there).
A simple and adequate  way to  specify  a string solution is
to write down the expression for  the string  action in the corresponding
background.
For example, a generalisation of \gu\
is represented by the folowing  conformal \sm
 (${\cal R}$ is 1/4 of the curvature of the  world-sheet metric)
\eqn\gene{ L_{PW}= \  (G_{\m\n} + B_{\m\n} )(x)  \  \del x^\m \bd x^\n  +
\a'{\cal R}\p (x) }
$$= \del u \bd v  +   K(x,u) \del u \bd u  +   2A_i (x,u) \del u
\bd x^i
+  \  \del x^i \bd x_i + \a'{\cal R}\p (u)  \ ,   $$
where the functions  $K,A_i,\p$ satisfy
\eqn\cone{   \del^i\del_i  K =  2\del^i\del_u A_i
+ 4 \del^2_u\p \ , \ \ \  \del_i { F}^{ij} = 0\  .  }
The $SO(D-3)$ invariant  ``plane wave" \gu\ with $K=K(r), \ r^2=x^ix_i$, i.e.
\eqn\hkh{ K ={ 1 + {2M \ov r^{D-4}} }   \ , \ \ D>4 \ ; \ \ \ K= { \  - {2M \
\ln \ {r\ov r_0} } } \ , \ \ \  D=4\ , }
is related (by space-time duality  in the $y$ direction \refs{\hhtt,\hhs})
to the following  ``fundamental string" (FS) solution \gibb\
\eqn\fs{ ds^2 = F(r) dudv + dr^2 + r^2 d\Omega^2_{D-3} \  ,  }
\eqn\bpf{B_{uv} = \ha F(r)\ ,\qquad  \e{2\p}  =  F(r) \  , \ \ \ F\equiv K\inv
\ ,  }
i.e. \horts\
\eqn\fss{   L_{FS} = F(r) \del u \bd v
+  \del x^i \bd x_i + \a'{\cal R}\p (r) \ .   }
This  background represents a  ``solitonic" string-like solution
of the ``vacuum" effective equations following from \act. At the same time,
it can  be considered   also as a   solution describing a
field outside an infinite  straight
fundamental string source along $y$ direction \gibb.
The string source provides the $\d$-function
in  the equation  for $F$:
$\  \del^i\del_i F\inv \sim \d (r)$.\foot{As  discussed in \refs{\tsmac,\tspa},
there exists a  (wormhole-type) resummation of the string perturbation theory
in which the  semiclassical approximation
is governed by the sum of the string action in a background of   massless
fields and the  classical effective action for the latter. FS can
then be considered as a solution corresponding to  such combined action. This
suggests that the need for a $\d$-function source  at $r=0$
is  related to  a non-perturbative solitonic nature
of the FS solution (note that the string action  and the effective action are
of different orders in string coupling).}

Fundamental string  solution is an extremal limit of a class of  leading-order
``black string" solutions  with axionic  charge \horstr\
(which can  be obtained by starting with a direct product of the  Schwarzschild
solution  and  a spatial direction and making a boost and duality
transformation
to generate an  axionic charge, i.e. a  non-trivial $B_{\m\n}$ \hhs).
FS is thus similar to  the extremal black holes
(with the role of the electric charge being played by the axionic one).

In contrast to non-extremal solutions (e.g. the neutral $D=4$ ``cosmic string"
field with a conical singularity)  which are modified by $\a'$ corrections,
the  FS   is exact (in a specially chosen scheme) \horts.
This is due to the special balance between the metric and the antisymmetric
tensor (and dilaton)
resulting in a very simple form of the corresponding string action \fss\
which, in particular,  has  conserved chiral currents (or additional
infinite-dimensional symmetries) and
is  conformally invariant.  This balance is also responsible for
the fact that the  equation for $F\inv$ is linear, i.e. that   FS solutions
with  parallel sources   can be linearly superposed  (this is also
the reason why  the force
on a static test string which is parallel to the string source vanishes
\refs{\gibb,\cakh}).

The FS solution has  a curvature singularity  at $r=0$ (and  no horizon)
and is stable and supersymmetric when embedded into a supergravity theory
\refs{\gibb,\wald}.  The effective string coupling
$\e{\p} $
 decreases to zero at  $r=0$  in   agreement with  the ``asymptotic freedom"
assumption, suggesting that
 the classical approximation can be trusted at small $r$.
In any case,  the  $r=0$ curvature singularity   does not necessarily mean
a singularity of  the string theory since  a propagation of test
quantum  strings in this background may not be  pathological.\foot{In fact,
scattering of test strings in this background appears to be
 similar to their scattering in flat space \refs{\cakh,\ghrw}.}

There are generalisations of the FS solution  (e.g. including  momentum
in the direction of the string and  Kaluza-Klein type charges).
Like \fs, they  are  leading-order solutions  \refs{\wald,\sen,\garf}
which turn out not to be  modified by $\a'$ corrections \hrts.
A particular  generalised  fundamental string    solution which we shall
consider  below  is    (cf. \fss)
\eqn\fsss{   L_{FS}' = F(r) \del u \bd v  +   b^2 \del u \bd u
+  \del x^i \bd x_i + \a'{\cal R}\p (r) \ ,   }
where $F$ and $\p$   are   the same as in  \hkh, \bpf.
 \fsss\ can be interpreted as
 describing  a field outside a closed   fundamental
 string winding   around  a  circular  spatial direction  in $u,v$ plane (of
radius
 $R_0= \sqrt{ \a'}/b$) \refs{\wald,\ghrw}.
The parameter $b$ can be set equal to 1 by  a rescaling of coordinates.
The string source
which is needed to support this  background at the origin  (i.e. to satisfy
the string  equations also at $r=0$)
is  represented by $x^i=0, \ u=2 R_0(\s +\tau), \ v= 2m (\s - \tau), \
m= R_0- \a' R_0\inv $, i.e. in addition to winding    there is a momentum  flow
along the string \ghrw.
All the main properties of the FS solution are true also for \fsss.
A novel feature is that the corresponding metric has a regular ergosphere \hrt.

\newsec{Extreme electric $ D=4$ black hole
as  dimensional reduction of exact $D=5$ fundamental string solution}
String theory naturally  leads to spaces of dimension $D> 4$.
If the scale  of ``extra" internal dimensions is small, non-trivial
higher-dimensional solutions may look like  $D=4$ ones.
To get  extremal black holes we need gauge fields. They may appear
as in  Kaluza-Klein model  if  internal dimensions are
compactified on
a torus.
{}From the  low energy field theory point of view we have
 to start with the full (containing terms of
all orders in  $\a'$)  massless
string effective action  in, say, five dimensions.
Assuming the fifth direction $y$ is periodic,  we can  expand the metric,
antisymmetric tensor and dilaton  in Fourier  series in $y$ and explicitly
integrate over $y$.
The result will be a $D=4$   effective action  containing massless fields
as well as  an  infinite tower
of massive modes with masses proportional to a  compactification scale.
Any exact solution of the $D=5$ theory which does not  depend on $y$ can then
be
directly interpreted as a solution of the equations of the
$D=4$ ``compactified" theory
with all massive modes set equal to zero (but all  ``massless" $\a'$-terms
included).
In string theory there is a simpler alternative -- to
 perform the dimensional reduction
 at the  more fundamental level
of the string action itself \hrts.   That amounts to
a separation of the action into gauge-invariant  $D=4$ part and
``internal" coordinate part, with the latter one  been coupled to gauge fields
originating from ($5\m$)-components of the metric and the antisymmetric tensor
(cf. \fgfg).

For example, consider the five dimensional plane fronted wave \gu,\hkh
\eqn\plwv{ ds^2_5 = du dv + K(r) du^2 + dr^2 + r^2 d\Omega^2_3\ , \ \ \ \ \ \
K(r)
= 1+{2M\ov r}\ .  }
 To reduce to four
dimensions  let us take  $u$  to be an internal coordinate $y$ and $v/2$ to be
the time coordinate $t$ in four dimensions.  Rewriting  \plwv\ in the form
\eqn\pwred{ ds^2_5 =  - K\inv {dt^2 } + dr^2 + r^2 d\Omega_2^2 + K ( dy +
{K\inv } dt)^2 \ ,  }
we conclude that  the four dimensional metric,  abelian gauge field and
the scalar $\s=\ha \ln G_{55}$ are
\eqn\exttt{ ds^2= - K\inv (r) dt^2 +  dr^2 + r^2 d\Omega^2_2 \ , \ \
\  A_t = K\inv (r)\ , \ \  \ \s =  {\ha} \ln K (r) \ . }
One thus finds
 the extremal electrically charged ``Kaluza-Klein" black hole
 which is a solution  corresponding to the action \actt\ with $a=\sqrt 3$ and
$\psi =\s$ \refs{\gib}. The $D=5$ solution  \plwv\ can be viewed  as a field
of a string source along $u$ direction boosted to the speed of light \gib.
At the same time, its  ``image" in the four dimensional space-time  orthogonal
to $u$-axis  is the extremal electric black hole.\foot{The same $D=4$ black
hole background is found
by dimensionally reducing the $D=5$ FS solution along $y=\ha (u+ v) $ direction
\gaun.
This is not surprising since the two (plane wave and FS) $D=5$ backgrounds are
related by duality. } Since \plwv\  is an exact string solution  we conclude
that \exttt\ is also  not  changed by $\a'$ corrections.

Starting with the $D=5$  generalised FS solution \fsss\
and dimensionally reducing to $D=4$ along $u$ direction
one finds \refs{\hrt,\hrts} the following $D=4$ background\foot{Dimensional
reduction along $u+v$ direction gives equivalent solution \hrts.}
\eqn\did{ ds^2 = - F^2 (r) dt^2 + dx_idx^i \ ,  } $$
\cA_t=-\cB_t = F(r)\  , \ \ \ \ \e{2\p}   = F (r)\ .  $$
Since the original $D=5$  background had a non-trivial antisymmetric tensor
we got $D=4$ solution with two vector fields  (cf. \fgfg).
The latter,  however,  are proportional to each other so \did\ can be
 identified  with the solution corresponding to \actt\ (with  $a=1,
\psi=2\phi$)
which describes the extremal electrically charged dilatonic black hole \elee.
The
 relation via dimensional reduction   to the exact
$D=5$ generalised FS solution implies that this extremal dilatonic black hole
is also an exact  solution of (bosonic as well as heterotic)
 string theory  \refs{\hrt, \hrts}.

\newsec{Conclusions}
It may be that string theory is ``asymptotically free", i.e.
 the  string coupling is small at small distances so that the dilaton
is massless and  the  theory is nearly  classical there.
 Then the dilaton  should play  an important role
in predictions of string theory about early Universe,
primordial black holes, etc.
The gravitational-type coupling of the dilaton
to gauge fields implies, in particular, that properties of the charged
black hole solutions are modified compared to  the pure Einstein-Maxwell
theory.

There exists a large class of classical string solutions with null Killing
vectors which do not receive  $\a'$ corrections, i.e. are exact.
Particular examples  are
 plane-wave type backgrounds and solitonic fundamental string solutions.
The effective string coupling (exponential of the dilaton)
 does indeed decrease  at small distances for the  latter solutions.

It turns out that these solutions are related  by dimensional reduction to
extremal
electric dilatonic black holes.
In particular, the $D=5$ solution describing the field
of a source represented by a closed fundamental string winding around
a compact 5-th dimension  appears  to be equivalent to the
$D=4$ extremal dilatonic black hole
with an electric charge of  Kaluza-Klein origin.
As a result, this  extremal black hole remains string solution to all orders in
 $\a'$.

There are various indications that extremal black holes   form
a class of ``solitonic" string solutions which are  very different from
non-extremal ones.
It appears that  non-extremal black holes
 (in particular, the Schwarzschild one) {\it are} modified by $\a'$ corrections
in an essential way.
A difficult unsolved   problem  is to find  an  exact classical string
solution  which generalises the Schwarzschild  solution to all orders
in $\a'$.
  It is usually conjectured  that it is quantum corrections that should be
responsible for elimination  of  black hole singularities.
However,  in non-local theories like string theory
 singularities may be absent already  at the classical level.
By  analogy with the open string case,  one may expect that  the modification
of the Schwarzschild  black hole
  may be similar to the one of  the  Coulomb point-charge solution in the
Born-Infeld  effective  theory:
the singularity  at the origin disappears being  ``regularised"  by the
string non-locality scale
$\sqrt{ \a'}$.


\newsec{Acknowledgement}
I would like to thank G. Horowitz and V. Mukhanov  for useful discussions and
acknowledge  also the  support of PPARC.


\listrefs
\end

{\footatend\immediate\closeout\rfile\writestoppt
\smallskip\newsec{ References}{\frenchspacing%
\parindent=20pt\escapechar=` \input refs.tmp\vfill\eject}\nonfrenchspacing}

\end